\documentclass[aps,prb,preprint,showpacs]{revtex4}
\usepackage{natbib}
\bibpunct{(}{)}{,}{a}{}{,}
\usepackage{epsfig}
\begin{document}
\title{Exact ground states for the four electron problem in a Hubbard ladder.}
\author{Endre~Kov\'acs and Zsolt~Gul\'acsi}  
\affiliation{Department of Theoretical Physics, University of Debrecen, 
H-4010 Debrecen, Hungary}
\date{\today}
\begin{abstract}
The exact ground state of four electrons in an arbitrary large two leg 
Hubbard ladder is deduced from nine analytic and explicit linear equations.
The used procedure is described, and the properties of the ground state are 
analyzed. The method is based on the construction in ${\bf r}$-space
of the different type of orthogonal basis wave vectors which span the 
subspace of
the Hilbert space containing the ground state. In order to do this, we start
from the possible microconfigurations of the four particles within the system.
These microconfigurations are then rotated, translated and spin-reversed in 
order to build up the basis vectors of the problem. A closed system of nine
analytic linear equations is obtained whose secular equation, by its minimum 
energy solution, provides the ground state energy and the ground state wave 
function of the model.
\end{abstract}
\maketitle

\section{Introduction}

Often, several authors have underlined that in understanding strong 
correlation effects presumably new methods of attack and new languages 
are necessary \citep{senthil}. Starting on this line, in the case 
of finite value of the concentration of carriers in the studied materials,
several non-perturbative procedures have been recently worked out, of which 
we mention the following ones. 
1) Bozonization techniques \citep{uu1} reactualized in mid-1990s
\citep{uu2,uu3} and even extended to finite temperatures \citep{uu4}, 
have been 
applied for several systems as Coulomb blockade in quantum dots \citep{uu5},
mesoscopic wires \citep{uu6}, two-band cases \citep{uu7}, ferromagnetism
\citep{uu8,uu9,uu10}, density waves \citep{uu11}, or metal-insulator 
transitions
\citep{uu12}. The procedure has been also used in the study of different model
characteristics for example, in the case of t-J \citep{uu13}, Luther-Emery 
\citep{uu14}, or Kondo \citep{uu15,uu16} models. 2) Infinite order canonical 
transformations applied for the Anderson and multiband Hubbard models
\citep{uu7,uu17}, and 3) Decomposition of the Hamiltonian in positive 
semidefinite operators, leading to interesting new phases \citep{uu18},
also in three dimensions \citep{uu19}, or even disordered and interacting 
\citep{uu20} cases. 

In the last years, several experimental results have directed the attention
to experimental situations suggesting interesting technological application 
possibilities, where the strong correlation effects emerge not at
finite concentration value of electrons, but for few electrons confined in a 
system or device. Such situations related to condensed matter physics, are 
encountered on a large spectrum of subfields, as for example in the case of
quantum dots \citep{i1}, quantum wells \citep{i2}, mesoscopics \citep{d5},
entanglement \citep{i5}, etc. This subject attracted increasing interest 
from the point of view of the theoretical description in the last decade.
Starting from even one electron problems solved exactly \citep{f1}, several 
cases of interest for two \citep{f2}, three \citep{i7a,i7b}, four \citep{d2}, 
or 
few \citep{f5a,f5b} particles have been studied.
Concretely, in the case of $N_p=4$ particles, even if the simulation have been
started more than a decade ago \citep{m02,m4}, only few valuable results are 
known in this subject in the condensed matter context, as for example the
energy dependence of the maximal Lyapunov exponent for 1D Lenard-Jones system 
\citep{d1}, the behaviour of a simplified spinless fermion correlated electrons
model \citep{d2}, or doped quantum well structures \citep{i2}.
As can be seen, the collected information, at least at the moment, provides 
only a poor characterization of the experimental situations of interest.

In this paper,
extending the frame of the non-perturbative methods mentioned above 
to the case of the low density limit, and starting from the aim to provide 
valuable high quality essential information for the $N_p=4$ case,
we provide exact results for this field, presenting the exact ground state 
for four interacting electrons in an arbitrary large two leg Hubbard ladder,
characterized by periodic boundary conditions. 
For this to be possible, a direct ${\bf r}$-space representation 
is used for the wave functions. On this line first symmetry adapted 
ortho-normalized basis wave vectors are constructed starting from local 
particle configurations. Based on these, an explicit and analytic closed 
system of equations is provided for nine type of basis wave vector components,
whose secular equation leads to the ground state wave function and the ground 
state energy of the system.

Deducing the ground state wave function for different microscopic parameters
of the model, ground state expectation values are calculated for different
physical quantities of interest, and correlation functions are deduced in 
order to characterize the ground state properties.  

For the case of the low density limit, the procedure presented here, 
enrols under the requirement of new languages. Compairing the deduced
ground states to ground states obtained in similar conditions for square
systems \citep{masikc} at exact level, the emerging differences suggest that
is highly questionable to approach at a good quality level, the two dimensional
behaviour from the ladder side.

The remaining part of the paper is structurated as follows. Section II. 
presents the Hamiltonian, Section III. describes the used procedure,
Section IV. characterizes the deduced ground states, and Section V. 
containing the summary and conclusions closes the presentation.

\section{The Hamiltonian}

The Hamiltonian of the Hubbard ladder is used in a standard form
\begin{eqnarray}
\hat{H}= t\hat{t} + U\hat{U},
\label{Eq1} 
\end{eqnarray}
where
\begin{eqnarray}
\hat{t}=\sum_{<i,j>,\sigma}(\hat{c}_{i\sigma}^{\dagger}\hat{c}_{j\sigma}+
\hat{c}_{j\sigma}^{\dagger}\hat{c}_{i\sigma}), \quad
\hat{U}=\sum^{N}_{j=1}\hat{n}_{j\uparrow}\hat{n}_{j\downarrow}\;.
\label{Eq2} 
\end{eqnarray}
In these expressions $\hat{c}_{i\sigma}$ are canonical Fermi operators which 
describe electrons on a two leg ladder, $N$ represents the number of lattice 
sites in the ladder, $\langle i,j \rangle$ denotes nearest-neighbour sites,
$t$ is the hopping matrix element for the electrons, and $U$ is the on-site
Coulomb repulsion.

Below we present the exact ground states of the presented model for $N_p=4$
particles and arbitrary ladder lenght, in the case of periodic boundary 
conditions and $u=U/t \geq 0$. The deduced ground state is a spin singlet state
and from physical point of view describes the repulsive Hubbard interaction 
case ($U > 0$), for $t> 0$, or the attractive Hubbard interaction case 
($U < 0$), for $t < 0$.

\section{The used procedure}

\subsection{The construction of the basis wave vectors}

The procedure we use is as follows. 1) First we number all lattice sites of the
ladder as shown in Fig.1, taking into account periodic boundary conditions,
and considering the number of rungs (e.g. $N/2$) integer number. 2) We 
identify the nine possible and qualitatively different microconfigurations 
in which four 
particles can be present into the ladder (see Fig.2). At this step we consider
that the ladder legs are equivalent, and spin reversed microconfigurations are
also equivalent. For the clarity of the mathematical notations we denote the
possible microconfigurations by capital letters $A$ to $J$, whose lower 
indices refer to the particle positions. In order to clearly distingwish the
microconfigurations, we must consider in the $C,E$ cases $i\ne j$, while in
the $F,J$ cases $j < k$ (see Fig.2). We further mention that in the process of
defining a microconfiguration, at least one of the particles is positioned
at site number $1$. Let us call these particles ,,starter'' particles of the
microconfiguration. For example, for the configuration $A_{i}$, the starter
particles are two electrons with opposite spin; for the microconfiguration
$F_{i,j,k}$ the starter particle is one electron with spin up, etc. 
(see Fig.2).  3) In the following step we define
nine different type of orthogonal basis wave functions connected to each 
possible microconfiguration, and denoted with the same capital letter 
introduced in a ket-vector. For example, connected to the microconfiguration
$A_{i}$, we have the basis wave vector $|A_{i}\rangle$; related to the
microconfiguration $B_{i}$ we have the basis wave vector $|B_{i}\rangle$, etc.
4) Since the lattice 
sites, legs, and spin orientations are equivalent, these properties must be
reflected also at the level of the basis vectors. Because of this reason, each
microconfiguration present in Fig.2 will be considered as generating 
microconfiguration and denoted by $O^{(1)}_{i,j,..}(n)$, where $n=1,2,...,9$
denotes the microconfiguration number.
For example $O^{(1)}_{i}(1)=A_{i}, O^{(1)}_{i}(2)=B_{i}, O^{(1)}_{i,j}(3)=
C_{i,j}, ...., O^{(1)}_{i,j,k}(9)=J_{i,j,k}$. 
5) Now, connected to each generating microconfiguration $O^{(1)}_{i,j,...}(n)$,
we define seven related (,,brother'') microconfigurations
$O^{(m)}_{i,j,...}(n)$ (e.g. $m=2,3,...8$) by a) rotating the generating
microconfiguration $m=1$ by 180 degree along the longitudinal symmetry
axis of the ladder, obtaining the $m=2$ brother microconfiguration, b)
rotating the generating microconfiguration $m=1$ by 180 degree along the 
symmetry axis perpendicular to the ladder, obtaining the $m=3$ brother
configuration, c) rotating the $m=3$ configuration by 180 degree along the
longitudinal symmetry axis of the ladder, obtaining the $m=4$ brother 
configuration, and finally d) reversing all spin orientations in the 
$m=1,2,3,4$ cases, we obtain the remaining $m=5,6,7,8$ brother 
microconfigurations. In this process it could happen that the starter
particles situated at site $1$ in the generating microconfiguration $m=1$,
arrive on the upper leg for a given brother configuration $m>1$.
An example of related (,,brother'') microconfigurations is presented in Fig.3.
for the $|C_{i,j}\rangle$ case.
6) At the sixth step, the effective construction of the basis vectors follows. 
A given basis vector $|O^{(1)}_{i,j,..} (n)\rangle$, of type $n$ ($n$ being
considered fixed), connected to the generating microconfiguration 
$O_{i,j,..}^{(1)}(n)$, is constructed as follows: a) The different
brother microconfigurations $O^{(m)}_{i,j,..}(n)$ are all translated 
along the ladder such that the starter particles (see point 1)), arrive on 
each lattice site of the same leg. In this process, the translation must be
such effectuated to not modify the interparticle positions 
and relative spin orientations inside the microconfiguration.
b) After this step, all obtained microconfigurations are added. 
c) The expression is written in mathematical form by representing each 
microconfiguration by four creation operators acting on the bare vacuum with 
no fermions present.
In order to do this, we have to fix the order of creation
operators for each basis vector type, which has been done as follows.
For two doubly occupied sites, we write the creation operators of
the couples next to each other, first the spin up, then the spin down 
contribution as
$\hat{c}^{\dagger}_{i\uparrow} \hat{c}^{\dagger}_{i\downarrow}
\hat{c}^{\dagger}_{j\uparrow} \hat{c}^{\dagger}_{j\downarrow} |0\rangle$.
In the case of basis vectors containing one doubly occupied site at $i$ we use
$\hat{c}^{\dagger}_{i\uparrow} \hat{c}^{\dagger}_{i\downarrow}
\hat{c}^{\dagger}_{j\uparrow} \hat{c}^{\dagger}_{k\downarrow} |0\rangle$.
Finally, for basis vectors without doubly-occupied sites, the
convention $\hat{c}^{\dagger}_{i\uparrow}\hat{c}^{\dagger}_{j\uparrow}
\hat{c}^{\dagger}_{k\downarrow}\hat{c}^{\dagger}_{l\downarrow} |0\rangle$
is considered, where $i<j$, $k<l$ must hold.

For example, using the above described steps for the $|C_{i,j}\rangle$
basis wave vector, one finds the eight related (,,brother'') 
microconfigurations 
as shown in Fig.3. Furthermore, the mathematical expression of 
$|C_{i,j}\rangle$ taken for example at $i=2, j=4$, becomes
\begin{eqnarray}
|C_{2,4}\rangle =
&&\Big( (\hat{c}_{1\uparrow}^{\dagger}\hat{c}_{1\downarrow}^{\dagger}
\hat{c}_{2\uparrow}^{\dagger}\hat{c}_{4\downarrow}^{\dagger}
+\hat{c}_{2\uparrow}^{\dagger}\hat{c}_{2\downarrow}^{\dagger}
\hat{c}_{3\uparrow}^{\dagger}\hat{c}_{5\downarrow}^{\dagger}+\dots)
\nonumber\\
&&+(\hat{c}_{(\frac{N}{2}+1)\uparrow}^{\dagger}\hat{c}_{(\frac{N}{2}+1)\downarrow}^{\dagger}
\hat{c}_{(\frac{N}{2}+2)\uparrow}^{\dagger}\hat{c}_{(\frac{N}{2}+4)\downarrow}^{\dagger}+
\hat{c}_{(\frac{N}{2}+2)\uparrow}^{\dagger}\hat{c}_{(\frac{N}{2}+2)\downarrow}^{\dagger}
\hat{c}_{(\frac{N}{2}+3)\uparrow}^{\dagger}\hat{c}_{(\frac{N}{2}+5)\downarrow}^{\dagger}+\dots)
\nonumber\\
&&+(\hat{c}_{4\uparrow}^{\dagger}\hat{c}_{4\downarrow}^{\dagger}
\hat{c}_{3\uparrow}^{\dagger}\hat{c}_{1\downarrow}^{\dagger}
+\hat{c}_{5\uparrow}^{\dagger}\hat{c}_{5\downarrow}^{\dagger}
\hat{c}_{4\uparrow}^{\dagger}\hat{c}_{2\downarrow}^{\dagger}+\dots)
\nonumber\\
&&+(\hat{c}_{(\frac{N}{2}+4)\uparrow}^{\dagger}\hat{c}_{(\frac{N}{2}+4)\downarrow}^{\dagger}
\hat{c}_{(\frac{N}{2}+3)\uparrow}^{\dagger}\hat{c}_{(\frac{N}{2}+1)\downarrow}^{\dagger}
+\hat{c}_{(\frac{N}{2}+5)\uparrow}^{\dagger}\hat{c}_{(\frac{N}{2}+5)\downarrow}^{\dagger}
\hat{c}_{(\frac{N}{2}+4)\uparrow}^{\dagger}\hat{c}_{(\frac{N}{2}+2)\downarrow}^{\dagger}+\dots)
\nonumber\\
&&+\hat{c}_{1\uparrow}^{\dagger}\hat{c}_{1\downarrow}^{\dagger}
\hat{c}_{4\uparrow}^{\dagger}\hat{c}_{2\downarrow}^{\dagger}
+\hat{c}_{2\uparrow}^{\dagger}\hat{c}_{2\downarrow}^{\dagger}
\hat{c}_{5\uparrow}^{\dagger}\hat{c}_{3\downarrow}^{\dagger}+\dots)+\dots \Big)
|0\rangle \: .
\label{Eq3}
\end{eqnarray}

We underline at this step, that because of the fixed conventions presented 
above, somethimes an additional negative sign arises in the process of 
writting the mathematical expressions corresponding to basis vector components
shifted from the end to the beginning of the ladder in the 
presence of periodic boundary conditions.
For example if we shift once more
$\hat{c}^{\dagger}_{1\uparrow}\hat{c}^{\dagger}_{N/2\uparrow}
\hat{c}^{\dagger}_{2\downarrow}\hat{c}^{\dagger}_{3\downarrow} |0\rangle$,
according to the fixed conventions one obtains
$\hat{c}^{\dagger}_{2\uparrow}\hat{c}^{\dagger}_{1\uparrow}
\hat{c}^{\dagger}_{3\downarrow}\hat{c}^{\dagger}_{4\downarrow} |0\rangle=
- \hat{c}^{\dagger}_{1\uparrow}\hat{c}^{\dagger}_{2\uparrow}
\hat{c}^{\dagger}_{3\downarrow}\hat{c}^{\dagger}_{4\downarrow} |0\rangle$.

\subsection{The deduction of the ground-state wave function}

Our basic observation leading to the solution of the problem is that
by applying the Hamiltonian $\hat H$ to a basis wave function 
$|O^{(1)}_{i,j,..}(n_1)\rangle$ holding a fixed $n=n_1$, only contributions
of the form $|O^{(1)}_{i,j,...}(n)\rangle$ with $n=1,2,...,9$ can be obtained 
as a result. Consequently nine (e.g. $n=1,2,..,9$) analytic equations 
building up a linear system of equations of the form
\begin{eqnarray}
\hat H |O^{(1)}_{i,j,..}(n)\rangle = \sum_{n'=1}^9 \, \sum_{i',j',..}
a^{n,n'}_{i',j',...} |O^{(1)}_{i',j',...}(n')\rangle,
\label{Eq4}
\end{eqnarray}
provide the solution of the problem for arbitrary ladder lenght (e.g. 
arbitrary $N$), where $a^{n,n'}_{i',j',..}$ are numerical coefficients. For 
example, in the case of $n=1$ (e.g. $|A_i\rangle=|O^{(1)}_{i}(1)\rangle$),
Eq.(\ref{Eq4}) becomes
\begin{eqnarray}
\hat H|A_{i}\rangle = 2 u |A_{i}\rangle - |D_{i,i}\rangle-\Theta(i>2)
|C_{i-1,i}\rangle - \Theta(i \leq N/4) |C_{i,i+1}\rangle,
\label{Eq5}
\end{eqnarray}
where $\Theta(K)=1$ if the condition $K$ is satisfied, otherwise $\Theta(K)=0$,
and we must have $1 < i \leq 1+N/4$. Analogously, for $n=2$ (e.g.
$|B_{i}\rangle=|O^{(1)}_{i,j,..}(2)\rangle$), Eq.(\ref{Eq4}) gives
\begin{eqnarray}
\hat H |B_{i}\rangle=2u|B_i \rangle - \Theta(i>1)|D_{i,i} \rangle 
- \Theta(i>1)|E_{i-1,i} \rangle
-\Theta(i \le N/4)|E_{i,i+1} \rangle,
\label{Eq6}
\end{eqnarray}
where for the index $i$ one must has $1 \leq i \leq 1+N/4$, 
and similar equations are obtained for the remaining $n=3,4,5,...,9$ 
index values as well. Consequently,
the nine orthogonal basis vectors presented in Fig.2. provide nine analytic
self-consistent linear equations (e.g. Eq.(\ref{Eq4})), containing the ground 
state of the problem. This means that solving the secular equation related to
Eq.(\ref{Eq4}) we arrive in a subspace of the original Hilbert space which 
contains the ground state. The ground state energy $E_g$ is the minimum 
possible energy provided by the mentioned secular equation, and the ground 
state wave function is the eigenvector corresponding to $E_g$. The ground 
state energy and the ground state wave function must be numerically obtained 
from the system of equations Eq.(\ref{Eq4}). The ground state nature of the
so obtained solution has been tested by numerical exact diagonalization taken
on the full Hilbert space at different $N$ values. 

The leading terms for two ground state wave functions 
deduced at a fixed $N$ and two different $u=U/t$ values are exemplified in 
the Appendix A. 

The fact that the solution of the ground state of four particles in an 
arbitrary 
large two leg ladder can be exactly given in Eq.(\ref{Eq4}), is related to the 
observation that the possible $n$ values (describing the different type of 
orthogonal basis vectors entering into the problem) are not changing if $N$ is
increased. Consequently, nine analytical equations will provide the solution in
Eq.(\ref{Eq4}) independent of how large the $N$ value is. But this does not 
mean that increasing $N$ in the course of the numerical treatement of 
Eq.(\ref{Eq4}), the same number of equations provided by Eq.(\ref{Eq4}) are 
encountered. This is because even if the analytic expression of 
$\hat H|O^{(1)}_{i,j,..}(n)\rangle$ at a fixed $n$ is the same for all $N$,
the domains covered by the indices $i,j,...$ in $|O^{(1)}_{i,j,..}(n)\rangle$
depend on the $N$ value. For example, the Eq.(\ref{Eq5}) represents the unique
analytic equation for the $|A_{i}\rangle$ base vector at arbitrary $N$. But
during the numerical treatement of Eq.(\ref{Eq5}), all equations for different
$i$ values must be considered. Since $i \leq 1+N/4$ holds, the unique analytic
equation (\ref{Eq5}), will provide an increasing number of different numerical
equations by increasing $N$. But even in this case, the Hilbert space region 
defined by Eq.(\ref{Eq4}) containing the ground state, has an accentuately
lower dimension $d_{red}$ than the dimension $d_H$ of the full Hilbert space 
of the problem. For example, at $N=16$ we have $d_H=14400$, $d_{red}=287$, 
while at $N=32$ one has $d_H=2.4 10^{5}$, but $d_{red}=2141$. As seen, at least
two orders of magnitude reduction in the number of basis wave vectors is 
encountered in the treatement of the problem.
 
At the level of principle, Eq.(\ref{Eq4}), based on symmetry properties, 
delimitates the region of the Hilbert space where the ground state is placed.
We must however emphasize that these symmetry properties, besides the 
symmetries of the system, depend on the microscopic parameters of the 
Hamiltonian as well. For example, Eq.(\ref{Eq4}) contains the ground state
wave function only for $u > 0$.

\section{The properties of the ground state}

Once the exact ground state wave function is known, the ground state itself 
can be characterized. Starting on this line, in this Section we exemplify the 
physical properties of the deduced ground state at $N=28$.

\subsection{Leading terms in the ground state, and ground state expectation 
values}

We analyze first the 
$u$ dependence of the leading terms of the ground state wave 
function $|\Psi_g\rangle$. As seen from Appendix A., at low $u$ the main 
contributions in $|\Psi_g\rangle$ are obtained from relatively closely situated
(e.g. almost nearest-neighbour) electrons with opposite spin (,,pairs''), the 
placement of the two pairs being such to maximize the interpair distance. 
Indeed, at $u=3$, the leading terms in Appendix A. are the $|D_{7,7}\rangle,
|E_{7,8}\rangle, |C_{7,8}\rangle, |D_{6,6}\rangle$ type of contributions,
which as seen from Fig.2. (taking into account that the described system is a
two leg ladder ring with 14 rungs), describe indeed this situation. As long as
$u$ is increased, the ,,pairs'' in the leading terms tend to form double 
occupancies  (as seen in Appendix. A. for $u=100$), the distance between the 
pairs remaining considerably high.

Now we turn to analyze ground state expectation values. The ground state 
expectation value of the kinetic energy term is presented in Fig.4. As seen, a
non-monotonic behaviour is obtained. The general tendency for the decrease
of the absolute value of $E_{kin}$ in function of $u$ at high $u$ can be 
understood by the double occupancy formation. The physical reason for the 
presence of the maximum in $E_{kin}$ for $u<15$ is not yet properly understood.
Probably the decrease in the increase rate of the double occupancy at a given
site above $u=5$ causes this behaviour (see Fig.5.).

The ground state expectation value of the double occupancy per site
$D = \langle \hat n_{i,\uparrow} \hat n_{i,\downarrow} \rangle$ is presented
in Fig.5. The general tendency present in this figure is that the
increasing $u$ value leads to the increase of $D$. This can be understood if we
remember that for $t<0$ the analyzed situation describes the attractive on-site
interaction case, hence by increasing $u$ we increase the number of double
occupancies. 

\subsection{Pair correlation functions}

Pair correlations are analyzed via the density-density correlation function
\begin{eqnarray}
C_n(r)=\frac{1}{N}\sum_{i=1}^N (\langle 
\hat{n}_i \hat{n}_{i+r} \rangle - 
\langle \hat{n}_i \rangle \langle \hat{n}_{i+r}\rangle ),
\label{corrn} 
\end{eqnarray}
and spin-spin correlation function defined by
\begin{eqnarray}
C_{S^z}(r)=\frac{1}{N}\sum_{i=1}^N (\langle 
\hat{S}^z_i \hat{S}^z_{i+r} \rangle 
- \langle \hat{S}^z_i \rangle \langle \hat{S}^z_{i+r}\rangle )=
 \frac{1}{2N}\sum_{i=1}^N (\langle 
\hat{n}_{i \uparrow}\hat{n}_{(i+r) \uparrow} \rangle - 
\langle \hat{n}_{i \uparrow}\hat{n}_{(i+r)\downarrow} \rangle )\:,
\label{corrs}
\end{eqnarray}
where $\hat{n}_i=\hat{n}_{i \uparrow}+\hat{n}_{i \downarrow}$, 
$\hat{n}_{i, \sigma}=\hat{c}_{i \sigma}^{\dagger}\hat{c}_{i \sigma}$,
$\hat S_i^{z}=(\hat n_{i,\uparrow}-\hat n_{i,\downarrow})/2$, $r$ measures
the inter-site distance in lattice constant units, and $\langle ...\rangle$
has the meaning of the ground state expectation value.

The behaviour of the spin-spin correlation function is presented in Fig.6. 
As seen, 
compaired to the non-interacting case, the spin-spin correlation decreases if 
$u$ is increased. Concerning the distance dependence of $C_{S^z}(r)$ at a 
fixed $u$, we see that even the short-range correlations are strongly 
(presumably exponentially) decreasing, and the spin correlation lenght covers
practically only nearest-neighbour sites.

The density-density correlations are exemplified in Fig.7. The presented 
curves have a specific structure which can be understood based on the analyzes
of the leading terms of the ground state wave function presented at the 
beginning of this Section. As observed there, in the ground state, two 
,,pairs'' tend to be situated at highest possible distance each from other,
providing the behaviour presented in Fig.7.

\section{Summary and conclusions}

We have deduced the exact ground state for four electrons in an arbitrary 
large two-leg Hubbard ladder and have analyzed the physical properties of the 
ground state in function of microscopic parameters of the model. 
The procedure is based on the construction in ${\bf r}$-space
of nine different type of orthogonal basis vectors which span the subspace of
the Hilbert space containing the ground state. In order to do this, we start
from the possible microconfigurations of the four particles within the system.
This microconfigurations are then rotated, translated and spin-reversed in 
order to build up the basis vectors of the problem. A closed system of linear 
equations is obtained whose secular equation, by its minimum energy solution, 
provides the ground state energy and the ground state wave function of the
model. The dimensionality of the subspace containing the ground state is
substantially less than the dimension of the full Hilbert space. The deduced
ground state wave functions have been used for the calculation of ground state
expectation values and correlation functions in the process of the 
characterization of ground state properties.

\acknowledgments

This work was supported by the Hungarian Scientific Research 
Fund through contract OTKA-T-037212. The numerical calculations have been 
done at the Supercomputing Lab. of the Faculty of Natural Sciences, 
Univ. of Debrecen, supported by OTKA-M-041537. 

\appendix

\section{Explicit ground state wave functions.}
\def\theequation{{\thesection}\arabic{equation}}
We present below the leading terms of explicit ground state wave functions 
deduced for $N=28$, at $u=3$ and $u=100$. The ground state $|\Psi_g\rangle$
is normalized to unity, and contains ortho-normalized basis vectors.

At $u=3$ we obtain for the ground state wave function
\begin{eqnarray}
&&|\Psi_g (u=3.0) \rangle = 
\nonumber\\
&&0.177614|D_{7,7}\rangle + 
0.170007|E_{7,8}\rangle + 
0.170006|C_{7,8}\rangle + 
0.15878|D_{6,6}\rangle \nonumber\\
&&+ 
0.158091|C_{6,7}\rangle +
 0.158089|E_{6,7}\rangle + 
0.138593|D_{8,7}\rangle + 
0.138593|D_{7,8}\rangle \nonumber\\
&&+ 
0.135254|C_{5,6}\rangle + 
0.135236|E_{5,6}\rangle + 
0.129199|D_{5,5}\rangle + 
0.128869|D_{6,7}\rangle \nonumber\\
&&+ 
0.128868|D_{7,6}\rangle + 
0.110223|D_{5,6}\rangle + 
0.110212|D_{6,5}\rangle + 
0.103555|C_{4,5}\rangle \nonumber\\
&&+ 
0.103502|E_{4,5}\rangle + 
0.100914|G_{7,8,1}\rangle -
 0.100914|G_{8,2,8}\rangle + 
0.0974412|E_{6,8}\rangle \nonumber\\
&&+ 
0.0974383|C_{6,8}\rangle -
 0.0938371|G_{7,2,7}\rangle + 
0.0938365|G_{6,7,1}\rangle + 
0.093035|D_{7,9}\rangle \nonumber\\
&&+ 
0.0917791|D_{4,4}\rangle + 
0.0897459|D_{8,6}\rangle + 
0.089745|D_{6,8}\rangle + 
0.0870033|C_{5,7}\rangle \nonumber\\
&&+ 
0.0869974|E_{5,7}\rangle + 
0.0843159|D_{4,5}\rangle + 
0.0842604|D_{5,4}\rangle -
 0.0822647|J_{7,1,8}\rangle \nonumber\\
&&+ 
0.082264|H_{7,7,14}\rangle -
 0.0802686|G_{6,2,6}\rangle + 
0.0802625|G_{5,6,1}\rangle + 
0.080119|D_{5,7}\rangle \nonumber\\
&&+ 
0.0801123|D_{7,5}\rangle + 
0.0774351|G_{7,8,14}\rangle -
 0.0771843|G_{8,2,9}\rangle + 
0.0764892|H_{6,6,14}\rangle \nonumber\\
&&-
 0.0764889|J_{6,1,7}\rangle + 
0.0746979|G_{6,7,14}\rangle -
 0.0746978|G_{8,2,7}\rangle + 
0.0744824|G_{7,8,2}\rangle \nonumber\\
&&-
 0.0744821|G_{7,2,8}\rangle + 
0.0704993|C_{4,6}\rangle + 
0.070455|E_{4,6}\rangle + 
0.0666869|G_{5,6,14}\rangle \nonumber\\
&&-
 0.0666861|G_{7,2,6}\rangle -
 0.0666109|G_{6,2,7}\rangle + 
0.0666052|G_{6,7,2}\rangle + 
0.0663597|C_{3,4}\rangle \nonumber\\
&&+ 
0.0663115|E_{3,4}\rangle + 
0.0654117|H_{5,5,14}\rangle -
 0.0654063|J_{5,1,6}\rangle + 
0.0650782|D_{8,8}\rangle \nonumber\\
&&+ 
0.064882|D_{4,6}\rangle + 
0.0648305|D_{6,4}\rangle -
 0.0614256|G_{5,2,5}\rangle + 
0.0613938|G_{4,5,1}\rangle \nonumber\\
&&+ 
0.0599305|G_{7,9,1}\rangle + 
0.0594696|E_{6,9}\rangle + 
0.0594586|C_{6,9}\rangle + 
0.0578123|G_{6,8,1}\rangle \nonumber\\
&&-
 0.0578118|G_{8,3,8}\rangle + 
0.0574775|D_{7,10}\rangle + 
0.0574745|D_{6,9}\rangle -
 0.0555399|H_{6,13,7}\rangle 
\nonumber\\
&&+ ...
\label{a1}
\end{eqnarray}
while for $u=100$ one has
\begin{eqnarray}
&&|\Psi_g (u=100)\rangle =
\nonumber\\
&&0.38466|B_7\rangle +
0.384611|A_7\rangle +
0.354728|B_6\rangle +
0.354558|A_6\rangle \nonumber\\
&&+
0.306686|B_5\rangle +
0.306065|A_5\rangle +
0.243462|B_4\rangle +
0.241179|A_4\rangle \nonumber\\
&&+
0.169981|B_3\rangle +
0.161576|A_3\rangle +
0.13959|B_8\rangle +
0.139581|A_8\rangle \nonumber\\
&&+
0.0961638|B_2\rangle +
0.0652468|A_2\rangle +
0.0623588|E_{7,8}\rangle +
0.0623528|C_{7,8}\rangle \nonumber\\
&&+
0.0615656|D_{7,7}\rangle +
0.0591513|E_{6,7}\rangle +
0.0591339|C_{6,7}\rangle +
0.0567649|D_{6,6}\rangle \nonumber\\
&&+
0.0529133|E_{5,6}\rangle +
0.0528502|C_{5,6}\rangle +
0.0490391|D_{5,5}\rangle +
0.044012|E_{4,5}\rangle \nonumber\\
&&+
0.0437798|C_{4,5}\rangle +
0.0387863|D_{4,4}\rangle +
0.0330753|E_{3,4}\rangle +
0.0322208|C_{3,4}\rangle \nonumber\\
&&+
0.0265348|D_{3,3}\rangle +
0.0223424|D_{8,8}\rangle +
0.0212907|E_{2,3}\rangle +
0.018145|C_{2,3}\rangle \nonumber\\
&&+
0.0171881|B_1\rangle +
0.0129086|D_{2,2}\rangle +
0.011569|E_{1,2}\rangle +
0.00249322|D_{8,7}\rangle \nonumber\\
&&+
0.00249322|D_{7,8}\rangle +
0.00236474|D_{6,7}\rangle +
0.00236474|D_{7,6}\rangle +
0.00211441|D_{5,6}\rangle \nonumber\\
&&+
0.00211441|D_{6,5}\rangle +
0.00175512|D_{4,5}\rangle +
0.00175512|D_{5,4}\rangle +
0.00130539|D_{3,4}\rangle \nonumber\\
&&+
0.00130539|D_{4,3}\rangle - 
0.0012476|G_{8,2,8}\rangle +
0.0012476|G_{7,8,1}\rangle +
0.00121559|E_{6,8}\rangle \nonumber\\
&&+
0.00121536|C_{6,8}\rangle +
0.00118331|G_{6,7,1}\rangle - 
0.00118331|G_{7,2,7}\rangle +
0.0011211|E_{5,7}\rangle \nonumber\\
&&+
0.00112029|C_{5,7}\rangle - 
0.00105805|G_{6,2,6}\rangle +
0.00105805|G_{5,6,1}\rangle +
0.000969643|E_{4,6}\rangle \nonumber\\
&&+
0.000966692|C_{4,6}\rangle - 
0.000878259|G_{5,2,5}\rangle +
0.000878259|G_{4,5,1}\rangle +
0.000788073|D_{2,3}\rangle \nonumber\\
&&+
0.00078803|D_{3,2}\rangle +
0.000771179|E_{3,5}\rangle +
0.000760318|C_{3,5}\rangle - 
0.000653214|G_{4,2,4}\rangle
\nonumber\\
&&+ ....
\label{a2} 
\end{eqnarray}



\newpage

\begin{figure}[h]
\centerline{\epsfbox{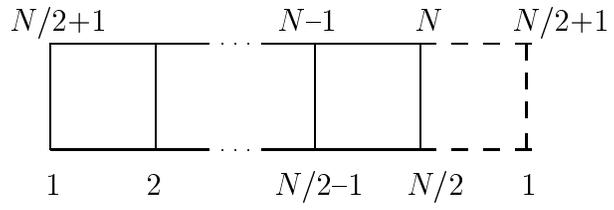}}
\caption{The numbering of the lattice sites for the two leg ladder
taken with periodic boundary conditions. $N$ is considered even.}
\label{fig1}
\end{figure}

\newpage

\begin{figure}[h]
\centerline{\epsfbox{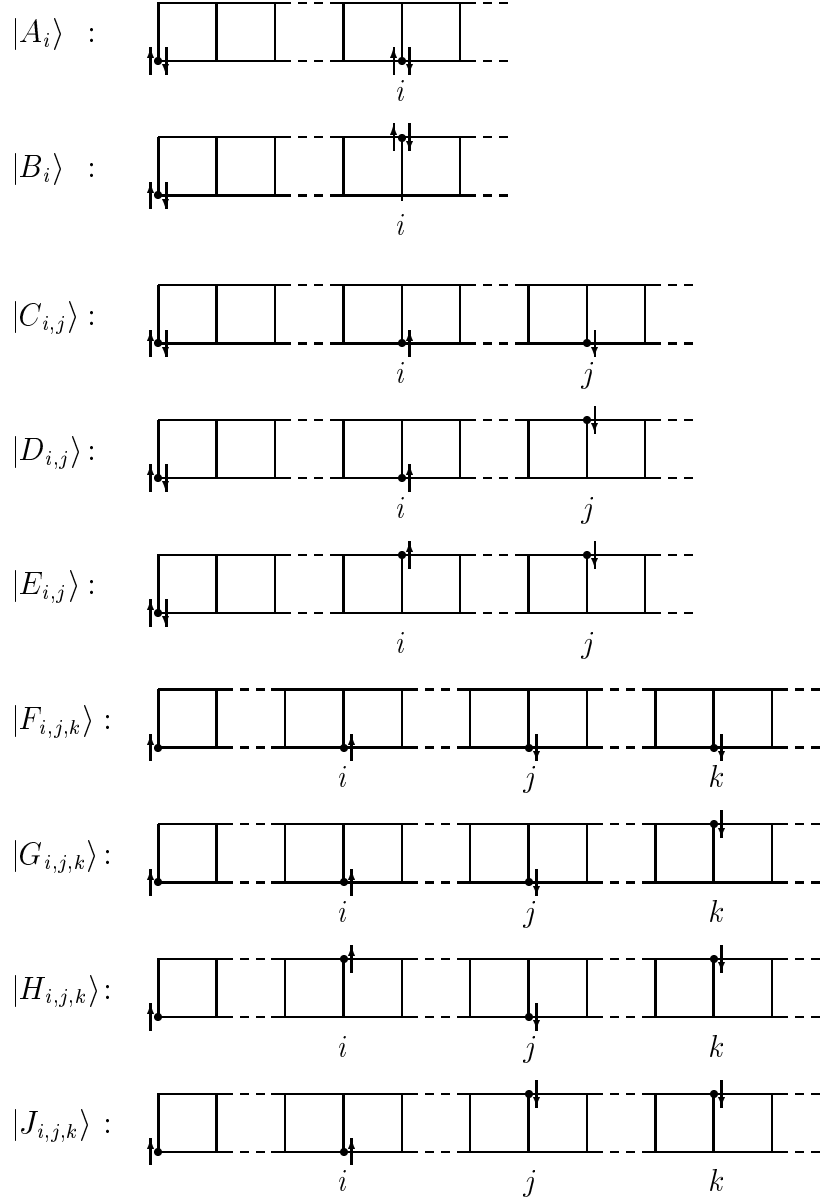}}
\caption{The different possible types of basis vectors. We note that for the 
cases $C,E$ $i\ne j$, while for $F,J$ $j<k$ is considered, respectively. In 
the cases $F,G,H,J$, the double occupancy is forbidden.}
\label{fig2}
\end{figure}

\newpage

\begin{figure}[h]
\centerline{\epsfbox{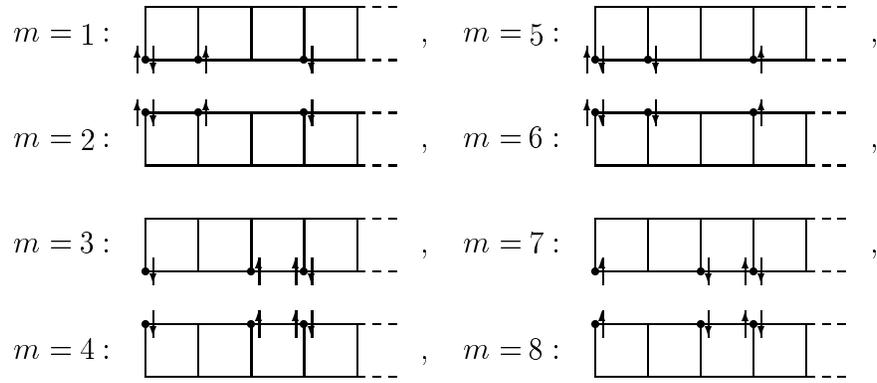}}
\caption{The brother microconfigurations together with their $m$ index
for the $|C_{i,j}\rangle$ basis wave vector.}
\label{fig3}
\end{figure}

\newpage

\begin{figure}[h]
\centerline{\epsfbox{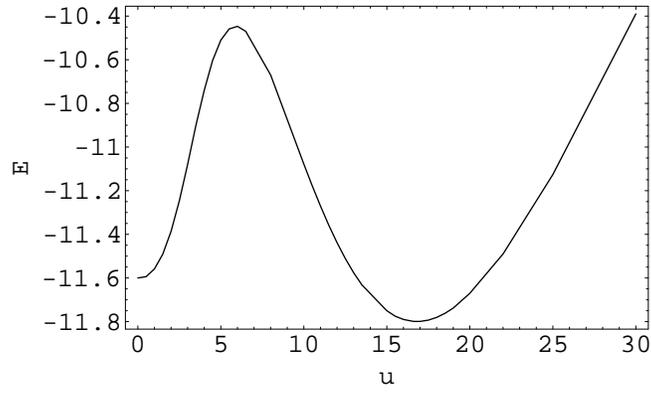}}
\caption{The dependence of the kinetic energy on $u$.}
\label{fig4}
\end{figure}

\newpage

\begin{figure}[h]
\centerline{\epsfbox{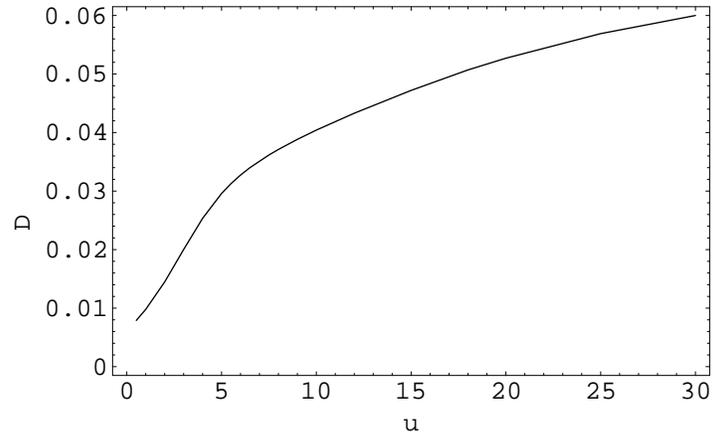}}
\caption{The dependence of the double occupancy per site 
$D = |\frac{E_{pot}}{uN}|$ on $u$.}
\label{fig5}
\end{figure}

\newpage

\begin{figure}[h]
\centerline{\epsfbox{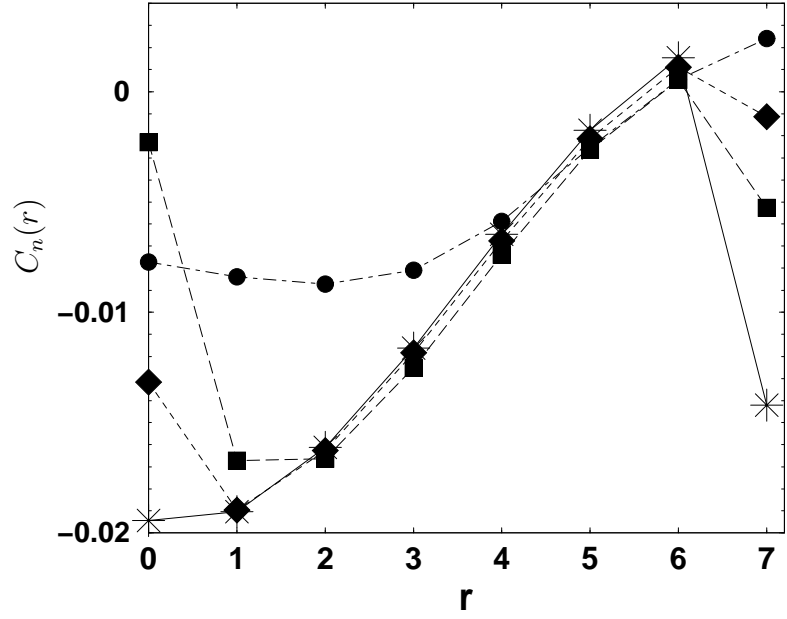}}
\caption{The density-density correlation function for 
$u=0$ (dots, dot-dashed line),
$u=10$ (squares, long dashed line),
$u=30$ (diamonds, short dashed line),
$u=100$ (stars, continuous line). $r$ is the distance in lattice 
constant units}
\label{fig6}
\end{figure}

\newpage

\begin{figure}[h]
\centerline{\epsfbox{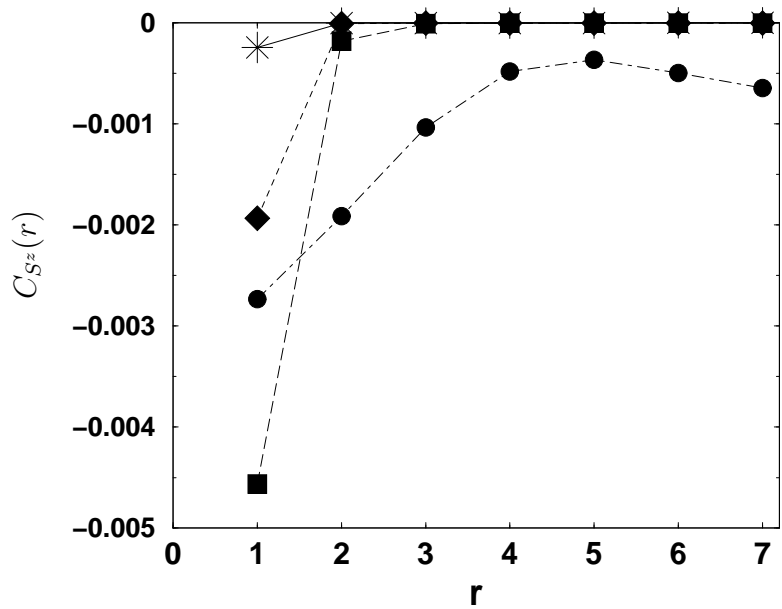}}
\caption{The $\hat{S}^z$-$\hat{S}^z$ correlation function for 
$u=0$ (dots, dot-dashed line),
$u=10$ (squares, long dashed line),
$u=30$ (diamonds, short dashed line),
$u=100$ (stars, continuous line). $r$ is the distance in lattice 
constant units}
\label{fig7}
\end{figure}

\end{document}